\documentclass[12pt]{article}
\usepackage{graphicx}
\usepackage[utf8]{inputenc}
\usepackage{amsmath}
\usepackage{amsfonts}
\usepackage{amssymb}
\usepackage{indentfirst}

\newcommand{\bfm}[1]{\mbox{\boldmath$#1$}}

\newcommand{\gsim}{\;\rlap{\lower 3.5 pt \hbox{$\mathchar \sim$}} \raise 1pt \hbox {$>$}\;}
\newcommand{\lsim}{\;\rlap{\lower 3.5 pt \hbox{$\mathchar \sim$}} \raise 1pt \hbox {$<$}\;}
\begin{document}

\title{
\vspace*{-20mm}
\begin{flushright}
{\normalsize ALBERTA-THY-11-17}\\[20mm]
\end{flushright}
\Large\bf
Coulomb Artifacts  and Breakdown of Perturbative Matching
in Lattice NRQCD}
\author{A.A. Penin and A. Rayyan\\[4mm]
{\it\normalsize Department of Physics, University of Alberta,} \\
{\it\normalsize  Edmonton, Alberta T6G 2J1, Canada}}
%\affiliation{Department of Physics, University of Alberta, Edmonton, Alberta T6G 2J1, Canada}
%\affiliation{Institut f\"ur Theoretische Teilchenphysik,
%Karlsruhe Institute of Technology, 76128 Karlsruhe, Germany}
%\affiliation{Institute for Nuclear Research, Russian Academy of Sciences, 117312 Moscow, Russia}
%\author{A. Rayyan}
%\affiliation{Department of Physics, University of Alberta, Edmonton, Alberta T6G 2J1, Canada}
%\date{}

%\preprint{ALBERTA-THY-??-17}
\date{}
\maketitle

\begin{abstract}
By studying an explicit analytical solution of the Schr\"odinger equation
with the Coulomb potential on the lattice we demonstrate a breakdown of
perturbative matching for the description of the Coulomb  artifacts in
lattice NRQCD, which leads to a large systematic error in the predictions for
the heavy quarkonium spectrum. The breakdown is a result of a fine interplay
between the short and long distance effects  specific to the lattice
regularization of  NRQCD.  We show how the problem can be solved by  matching
the lattice and continuum results for the solution of the full  Schr\"odinger
equation without the expansion in the Coulomb interaction.
\end{abstract}

%\pacs{ 12.38.Gc, 12.38.Bx, 14.40.Pq, 14.65.Fy}

\section{Introduction}
\label{sec::int}
The lattice simulations within the effective theory of nonrelativistic QCD \linebreak
(NRQCD) \cite{Caswell:1985ui,Bodwin:1994jh,Thacker:1990bm,Lepage:1992tx} has
developed into one of the most powerful tools for the theoretical analysis of
heavy quarkonium properties \cite{Dowdall:2011wh}. This method is entirely based
on first principles, allows for  simultaneous treatment of dynamical heavy and
light quarks, and gives a systematic account of the long distance nonperturbative
effects of the strong interaction. A crucial part  of the  approach is the
matching of lattice NRQCD to the  full theory of  continuum  QCD, which properly
takes into account the effect of the hard relativistic modes. For a long time
this   procedure was thought to be well understood. Recently, however, a problem
of the standard perturbative matching  in the analysis of the heavy
quark-antiquark bound states on the lattice has been pointed out
\cite{Liu:2016fus}. The problem  is related to the description of the lattice
Coulomb artifacts  resulting from the effect of the space discretization  on the
Coulomb bound state dynamics. The Coulomb artifacts appear as powers of a
dimensionless combination $\alpha_s a m_q$ of the strong coupling  constant
$\alpha_s$, lattice spacing  $a$, and the heavy quark mass $m_q$ in the
parameters of the  bound states evaluated on the lattice. This dependence on the
lattice spacing should be  cancelled in the final result for the physical
quarkonium spectrum through the matching procedure and an inconsistent treatment
of the artifacts  may lead to  a large systematic error of the lattice NRQCD
predictions.  In the present paper, by studying an explicit  analytical solution
of the Schr\"odinger equation on the lattice, we show that the standard
finite-order perturbative matching  is  insufficient in dealing with the Coulomb
artifacts, confirming the result of numerical analysis \cite{Liu:2016fus}. This
appears  counterintuitive since the lattice regularization is usually associated
with a momentum cutoff at the scale $1/a$ much larger than the scale $\alpha_s
m_q$ of Coulomb dynamics and  the corresponding short-distance effects are
supposed to be systematically described by the standard matching procedure.  We
find  that the failure  of perturbative matching is a consequence of a fine
interplay between the short and long distance effects  specific to the lattice
regularization of NRQCD. The paper is organized as follows. In the next section
we consider the case of the four-quark operators in the NRQCD Lagrangian
where the perturbative matching breaks down in one loop, and show how the
problem can be related to the properties of the Coulomb wave function on the
lattice within the method of ``Schr\"odinger  matching''. In Sects.~\ref{sec::3}
and \ref{sec::4} we describe an  exact analytical solution  of the Schr\"odinger
equation with Coulomb potential on a spatial lattice, which is used to  model
the  Coulomb artifacts of the real lattice NRQCD simulations without the
expansion in $\alpha_s$, and explain the mechanism of the perturbative matching
breakdown. In Sect.~\ref{sec::5} we  discuss the application of our result to
the  analysis of the  bottomonium spectrum in lattice NRQCD.

\section{Perturbative and Schr\"odinger matching of four-quark operators}
\label{sec::2}
\begin{figure}[t]
\begin{center}
\begin{tabular}{c}
%\hspace*{6mm}
\includegraphics[width=5.0cm]{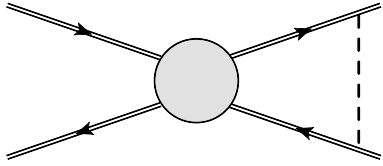}\\[5mm]
\end{tabular}
\end{center}
\caption{\label{fig::fig1}  One-loop Feynman diagram contributing to the ${\cal
O}(v^4)$   one-particle irreducible scattering amplitude in  NRQCD.  The double
arrow and  dashed  lines stand for the nonrelativistic quark and  Coulomb gluon
propagators, respectively. The gray circle denotes the effective  ${\cal
O}(v^4)$ interaction generated by a single gluon exchange. The symmetric diagram is
not shown. }
\end{figure}

The effect under consideration is characteristic to the four-quark operators
contribution to the NRQCD Lagrangian
\begin{equation}
\delta{\cal L}_{NRQCD}=\sum_i{C_F\alpha_s\over m_q^2}C_iO_i\,,
\label{eq::4quark}
\end{equation}
where
\begin{equation}
O_i=\psi^\dagger {\bfm \Gamma}_i\psi \chi_c^\dagger
{\bfm \Gamma}_i\chi_c\,,
\label{eq::Oi}
\end{equation}
$\psi$ ($\chi_c$) are the nonrelativistic Pauli spinors of quark (antiquark)
field, ${\bfm \Gamma}_i$ is a matrix  in color and spinor space and $C_i$ is the
corresponding   Wilson coefficient.  The coefficients $C_i$ vanish in the Born
approximation and is determined order by order in $\alpha_s$ by equating the
one-particle irreducible quark-antiquark  scattering amplitudes in QCD  and
NRQCD in a given  order in heavy quark velocity $v$. The Coulomb artifacts are
associated with the static Coulomb interaction between  the  nonrelativistic
quark and antiquark. The  nonrelativistic  Coulomb dynamics is  not sensitive to
the high momentum region and, to the leading order in $v$, the result for such a
contribution obtained within continuum QCD and NRQCD  coincide. Thus the
matching is determined by the difference between continuum and lattice NRQCD
amplitudes. In the one-loop approximation the relevant NRQCD diagram is given in
Fig.~\ref{fig::fig1}, where the gray circle  represents the  effective ${\cal
O}(v^4)$ interaction generated by a  tree gluon  exchange.\footnote{The leading
order NRQCD Lagrangian is ${\cal O}(v^2)$.} It,  in particular, includes the
contact Fermi and  Darwin terms proportional to the delta-function of the
distance between the heavy quark and antiquark, which have the structure of
Eq.~(\ref{eq::Oi}). These terms  determine the ${\cal O}(\alpha^2_s)$
perturbative corrections to the Coulomb energy levels given, up to an overall
factor, by  the matrix element $\langle O_i\rangle$ of the corresponding
operators  between the quarkonium states, which is proportional to the square of
the  wave function at the origin  $|\psi(0)|^2$.

Within the  standard perturbative matching  the difference of the above diagram
computed in the continuum  and on the lattice should be included into the
one-loop  Wilson coefficients $C_i$.  The corresponding corrections  to the
energy level are proportional to $C_i\langle O_i\rangle$. On the other hand the
diagram in  Fig.~\ref{fig::fig1} is quite special since it contains a static
Coulomb exchange, which is responsible for the formation of the Coulomb-like
heavy quarkonium states.  When sandwiched between the Coulomb states such an
exchange can be absorbed into the Coulomb wave function by  the equation of
motion, {\it i.e.} the diagram is already contained in the Coulomb matrix
element of the ${\cal O}(v^4)$ tree gluon exchange. Thus the matching for this
diagram as well as for the diagrams with any number of the Coulomb exchanges can
be incorporated into the ``Schr\"odinger matching'' for the value of the wave
function at the origin introduced in Ref.~\cite{Liu:2016fus}. The corresponding
correction to the energy level is proportional to
\begin{equation}
\left(1-{|\psi_l(0)|^2\over |\psi_c(0)|^2}\right)\langle O_i\rangle\,,
\label{eq::Schmatch}
\end{equation}
where $\psi_c$ ($\psi_l$) stands for the Coulomb wave function computed in the
continuum (on the lattice). One may suggest that for $\alpha_s m_q\ll 1/a$ the
two procedures should give equivalent results and the Coulomb artifact
contribution to $C_i$ should reproduce Eq.~(\ref{eq::Schmatch}) in a given order
of the expansion in $\alpha_s$. However in Ref.~\cite{Liu:2016fus} it has been
demonstrated by numerical  analysis of the Schr\"odinger equation  that the
effects of space discretization  on the Coulomb dynamics  {\em cannot} be
accounted for order by order in $\alpha_s$ and require the exact solution of the
Coulomb problem without the expansion in strong coupling constant. In
particular the absence of the linear Coulomb  artifacts predicted by the
one-loop  matching has been observed. At the same time a rather subtle
mechanism of the perturbative matching breakdown cannot be  easily traced within
a numerical approach. In the next sections we study an analytic solution of the
Coulomb problem on the lattice and give a detailed  explanation of this
mechanism.

\section{Analytical solution of the Coulomb problem  on the lattice}
\label{sec::3}
We consider a stationary Schr\"odinger  equation with the Coulomb potential
$V(r)=-C_F\alpha_s/r$ on a spatial lattice and restrict the analysis to the
spherically symmetric $S$-states so that only the discretization of the radial
coordinate $r=an$, $n=0,1,\ldots$ is necessary. The time variable is kept
continuous. This simplified model retains the main features of the  real lattice
NRQCD simulations discussed in the next section.  For the central-difference
discretization of the radial derivative the Schr\"odinger equation for the
function  $R(n)=r \psi_l(n)$  becomes
\begin{equation}
R(n+1)- 2 R(n) + R(n-1) + 2a^2 \left(E+ \frac{1}{na}\right)R(n) = 0\,,
\label{eq::diffeq}
\end{equation}
with the boundary condition $R(0)=0$.  Throughout  this section we use the
Coulomb units, with the reduced Bohr radius $r_B=(C_F\alpha_sm_q/2)^{-1}$ being
the unit of length.  The analytical solution of the eigenstate problem
Eq.~(\ref{eq::diffeq}) is known and the corresponding eigenfunctions for
discrete and continuum spectrum are found in terms of a hypergeometric function
\cite{Kvi:1992}.  For  the ground state the solution takes a particulary simple
form
\begin{eqnarray}
E_l &=& -\frac{1}{a^2} \left((1+a^2)^{1/2} - 1 \right) =  E_c\left(1 -{a^2\over 4}
+ {\cal O}(a^4)\right),
\label{eq::El} \\
\psi_l(n) &=&  \frac{1}{\pi^{1/2}(1+a^2)^{1/4}}
e^{-n\,{\rm sinh}^{-1}a} = \psi_c(r)+ {\cal O}(a^2)\,,
\label{eq::psil}
\end{eqnarray}
where $E_c=-\frac{1}{2}$ and $\psi_c(r)=\frac{1}{\sqrt{\pi}}\, e^{-r}$
are the well known  continuum results for the ground state energy and wave
function,  and the normalization condition reads
\begin{equation}
4\pi a^3 \sum_{n=0}^{\infty} n^2|\psi_l(n)|^2 = 1\,.
\end{equation}
The wave function in the momentum space is defined
by the Fourier transform
\begin{equation}
\tilde{\psi}_l(p)
%= a^3 \int {\rm d}\Omega\sum_{n=0}^\infty  n^2 \psi_l(n)\,e^{-i n a p \text{ cos}( \theta)}
= 4 \pi a^3 \sum_{n=0}^\infty n^2 \psi_l(n) \frac{\sin(nap)}{nap}
\label{eq::FT}
\end{equation}
and reads
\begin{equation}
\tilde{\psi}_l(p) = \frac{ 8\sqrt{\pi}}{(1+a^2)^{1/4}}
\frac{(a/2)^4}{\left(\sin^2(pa/2)
- a^2 E_l/2\right)^2} \frac{\sin(ap)}{a p}\\
= \tilde\psi_c(p) + {\cal O}(a^2)\,,
\label{eq::psilp}
\end{equation}
where $\tilde\psi_c(p)=\frac{8 \sqrt{\pi}}{(p^2 + 1)^2}$ is the continuum
result in Coulomb units. A peculiar feature of the  Fourier transform
is that $\psi_l(0)$ does not contribute to Eq.~(\ref{eq::FT}) due to the $n^2$
factor. As a consequence the inverse  Fourier transform
\begin{equation}
\psi_l(n)= \frac{1}{2 \pi^2} \int_0^{\frac{\pi}{a}}
{\rm d}p\, p^2 \,\tilde{\psi}_l(p)\, \frac{\sin(nap)}{nap}
\label{eq::invFT}
\end{equation}
give the correct value of $\psi_l(n)$ only for the sites with $n\ne 0$. For
$n=0$ the  corresponding integral\footnote{The subscript $p$ in the
Eq.~(\ref{eq::psip0int}) is used to distinguish the formal result of the
inverse Fourier transform from the actual value of the solution
Eq.~(\ref{eq::psil}) at $n=0$.}
\begin{equation}
\psi_{p}(0)\equiv\frac{1}{2 \pi^2} \int_0^{\frac{\pi}{a}}
{\rm d}p\, p^2 \,\tilde{\psi}_l(p)
\label{eq::psip0int}
\end{equation}
takes the value
\begin{equation}
\psi_{p}(0)=
\frac{1}{\sqrt{\pi}}\left(1-{\sqrt{1+a^2}-1\over a}\right)
=\psi_c(0)\left(1-{a\over 2}+{\cal O}(a^2)\right),
\label{eq::psip0}
\end{equation}
while the original solution in the coordinate space is
\begin{equation}
\psi_l(0)= \frac{1}{\sqrt{\pi}(1+a^2)^{1/4}}= \psi_c(0)
\left(1-{a^2\over 4}+{\cal O}(a^4)\right),
\label{eq::psil0}
\end{equation}
in full agreement with the  results of the numerical analysis
\cite{Liu:2016fus}.  Though both  values have the same continuum limit, they
approach it at a different  rate since   Eq.~(\ref{eq::psip0}) has a
nonvanishing ${\cal O}(a)$ term. In the Sect.~\ref{sec::5} we show this to be
precisely the origin of the spurious linear artifact in the result of
perturbative matching. We should emphasize that Eq.~(\ref{eq::psip0}) defines a
function,  which  does not satisfy the  Schr\"odinger  equation in the limit
$a\to 0$. Indeed, let us consider the (forward) derivative of the wave function
at the origin.  Then for the solution  Eq.~(\ref{eq::psil}) we get
\begin{equation}
\psi'_l(0)= {\psi_l(1)-\psi_l(0)\over a}=-\psi_c(0)+{\cal O}(a)\,,
\label{eq::psiprl}
\end{equation}
which in the limit $a\to 0$ recovers the  property of the continuum solution
\begin{equation}
\psi_c'(0)=-\psi_c(0)\,.
\label{eq::rel}
\end{equation}
At the same time if Eq.~(\ref{eq::psip0}) is used to define
the value of the wave function at the origin we get
\begin{equation}
\psi'_{p}(0)= {\psi_l(1)-\psi_{p}(0)\over a}
=-{1\over 2}\psi_c(0) +{\cal O}(a)\,,
\label{eq::psiprp}
\end{equation}
which violates the continuum result at ${\cal O}(1)$.

Finally we should note that while the linear term in Eq.~(\ref{eq::psip0}) is
universal for all the $S$-wave states, the coefficient of the quadratic terms
in the expansion of the bound state parameters in $a$ is sensitive to the
Coulomb dynamics, {\it e.g.} for the  first excited state this coefficient in
Eq.~(\ref{eq::psil0}) changes from $-1/4$ to $1/16$.

\section{Breakdown of perturbative matching}
\label{sec::4}
Let us now consider the Coulomb contribution of the   diagram
Fig.~\ref{fig::fig1} to a Wilson coefficient $C_i$. As it has been  pointed out
this contribution  is given by the difference of the continuum and lattice NRQCD
one-loop expressions and can be computed in the kinematics where the heavy
quarks are at rest and on the threshold. In the given order of the
nonrelativistic expansion  it reads
\begin{equation}
\delta C_i= \frac{4C_F\alpha_s}{\pi } \left(\int_\lambda^{\infty}
{\rm d}p\, p^2  D_c(p)G_c(p)- \int_\lambda^{\frac{\pi}{a}}
{\rm d}p\, p^2  D_l(p)G_l(p)\right),
\label{eq::delCint}
\end{equation}
where the integration over the time component of the virtual momentum has
been performed by taking the residue of the  heavy quark propagator, an
auxiliary infrared cutoff  $\lambda$ is introduced to regularize the
integrals at small momentum,  and we switch back  to the standard units. The
continuum Coulomb gluon and heavy quark propagators
\begin{equation}
D_c(p)={1\over p^2}\,,\qquad
G_c(p)={m_q\over p^2}\,,
\label{eq::contprop}
\end{equation}
correspond to the leading order NRQCD  action.  Their  lattice counterparts
\begin{equation}
D_l(p)=\frac{\sin(ap)}{a p}{(a/2)^2\over \sin^2(ap/2)}\,, \qquad
G_l(p)={m_q(a/2)^2\over \sin^2(ap/2)}\,
\label{eq::latprop}
\end{equation}
exactly correspond to the discretization of the  Schr\"odinger
equation~(\ref{eq::diffeq}). After integration we get
\begin{equation}
\delta C_i={1\over 2}C_F\alpha_s a m_q+{\cal O}\left(\left(\lambda a\right)^2\right)\,.
\label{eq::delC}
\end{equation}
Numerically, the main effect of the lattice regularization comes
from the ultraviolet momentum cutoff at $p=\pi/a$. Indeed, if the continuum
propagators are used  in the second term of Eq.~(\ref{eq::delCint}) the
coefficient in Eq.~(\ref{eq::delC}) is changed only from $1/2$ to
$4/\pi^2=0.405\ldots$.

The lattice contribution to Eq.~(\ref{eq::delCint}) can  be obtained by the
expansion of $\tilde\psi_l(p)$ in Eq.~(\ref{eq::psip0int}) to the first order
in $\alpha_s$.\footnote{In the Coulomb units this is equivalent to the
expansion in $a$ with fixed   $ap\sim 1$.} Hence, as one may expect from the
general arguments, Eq.~(\ref{eq::delC}) agrees with the expansion of the
first factor in Eq.~(\ref{eq::Schmatch}) to the first order in $\alpha_s$ if
the momentum space result Eq.~(\ref{eq::psip0}) is used to define the value
of the wave function at the origin. However Eq.~(\ref{eq::psip0}) does not
agree with the actual solution Eq.~(\ref{eq::psil0}) and results in  a
pathological wave function which does not satisfy the  Schr\"odinger equation
in the continuum limit at $r=0$. Thus the Wilson coefficient
Eq.~(\ref{eq::delC}) {\it does not} cancel the dependence of the ${\cal
O}(v^4)$ tree gluon exchange matrix element on $a$ and  we observe a
breakdown of  perturbative matching in the analysis  of the Coulomb artifacts
already in one loop.  By contrast the  Schr\"odinger matching with the exact
solution of the Coulomb problem on the lattice gives the correct result for
the Coulomb artifacts to all orders in $\alpha_s$.

We can trace the origin  of this phenomenon to the fact that the relation
Eq.~(\ref{eq::rel}) violated by Eqs.~(\ref{eq::psip0},\,\ref{eq::delC})
follows from the cancellation of the singular kinetic and potential energy
terms in the  Schr\"odinger equation at  $r \to 0$. Indeed in the  continuum
the radial equation for the $S$-wave function reads
\begin{equation}
\left({d^2\over dr^2}+{2\over r}{d\over dr}
-{C_F\alpha_sm_q\over r}+m_qE\right)\psi(r)=0\,.
\label{eq::conteq}
\end{equation}
Keeping the most singular terms in the limit  $r \to 0$ we get
\begin{equation}
\left.\left({d\over dr}-{1\over r_B}\right)\psi(r)\right|_{r\to 0}=0
\label{eq::limit}
\end{equation}
which reproduces Eq.~(\ref{eq::rel}) in Coulomb units. The first (second)
term in Eq.~(\ref{eq::limit})  corresponds to the kinetic (potential) energy
contribution.  Hence at the origin these terms  should be considered on an
equal footing while the standard matching treats the Coulomb potential as a
perturbation. Evidently the above mechanism of the perturbative matching
breakdown is specific to the lattice regularization and the contact
interaction in the NRQCD Lagrangian. We can regularize  the contact
interaction by separating the quark and antiquark fields with a small spatial
interval $r_0\sim a$ and take the limit $r_0\to 0$ after the matching is
done. This introduces an additional $\sin(r_0p)/(r_0p)$ factor into the
integrands of Eq.~(\ref{eq::delCint}). The lattice integral then reads
\begin{equation}
m_q\int_\lambda^{\frac{\pi}{a}}
{\rm d}p\, p^2 {\sin(r_0p)\over r_0p}\frac{\sin(ap)}{a p}
\left({(a/2)\over \sin(ap/2)}\right)^4=
m_q\left({1\over \lambda}-{\pi\over 4} r_0\right)
+{\cal O}\left((\lambda a)^2\right)\,,
\label{eq::r0}
\end{equation}
which removes  ${\cal O}(a)$ contribution from the  Wilson coefficient
Eq.~(\ref{eq::delC}) and brings it into agreement with the Schr\"odinger
matching result.

We should emphasize the difference between matching calculations in lattice
NRQCD and in NRQCD   with an explicit momentum cutoff $\Lambda_{UV}\sim1/a$,
which is not plagued with the problem discussed above. In the latter theory
the properties of the solution of the Coulomb  problem in the (continuum)
coordinate space are significantly different from  the solution of the finite
difference equation Eq.~(\ref{eq::diffeq}).  The Schr\"odinger equation in
this case is a differential equation and its regular solution satisfies the
conditions of the Fourier inversion theorem. Hence the value of the wave
function at the origin is unambiguously determined by the integral of the
wave function in momentum space and the problem discussed in the previous
section does not exist. The correct behavior of $\psi(r)$ and $\psi'(r)$ at
$r\to 0$ then follows from the continuity and smoothness of the solution.  A
comprehensive analysis of the four-fermion operator matching  with an
explicit momentum cutoff can be found in Ref.~\cite{Hill:2000qi} in a context
of the NRQED calculation of the radiative corrections to the orthopositronium
decay rate. In \cite{Hill:2000qi} a numerical solution of the
coordinate-space Schr\"odinger equation has been obtained for the Hamiltonian
defined in a momentum cutoff regularization scheme.  It has been found that
the dependence of the value of the resulting wave function at the origin on
the cutoff includes  a linear  Coulomb artifact  which is cancelled  by the
${\cal O}(\alpha_sm_q/\Lambda_{UV})$  term  in the one-loop Wilson
coefficient. Thus in the momentum cutoff scheme the results of the
perturbative and Schr\"odinger matching do agree.

The discretization method and the action of the model discussed above do not
exactly reproduce  the ones used in the real lattice simulations. The inclusion
of the relativistic corrections, a different discretization of the gluon field
or the use of a cubic lattice would change the  dependence of $\psi_l(0)$ and
$C_i$ on $a$.  However the Wilson coefficient obtained within the perturbative
matching  always has a linear dependence on $a$ from the momentum cutoff in
Eq.~(\ref{eq::delCint}). At the same time the solution of the Schr\"odinger
equation with the central difference discretization of the kinetic energy
operator has only  ${\cal O}(a^2)$ global error (see {\it e.g.}
\cite{Hairer:2008}) and is free of the linear artifacts. This is confirmed by a
numerical analysis of the discretized Schr\"odinger-Pauli equation at ${\cal
O}(v^4)$ on a cubic lattice \cite{Bali:1998pi}, which with a good precision rules
out the linear dependence of the bound state parameters on the lattice spacing.
In general the Schr\"odinger matching can be performed numerically for any
given Hamiltonian and the  lattice used in the simulations, but even a simple
analysis based on Eq.~(\ref{eq::psil0}) gives a  good estimate of the dependence
of the bare lattice NRQCD data for the hyperfine splitting in bottomonium on $a$
\cite{Liu:2016fus}.

\section{Coulomb artifacts and  bottomonium  spectrum
in  lattice  NRQCD}
\label{sec::5}
Let us now consider how the above analysis affects the determination of the
bottomonium  spectrum in  the radiatively improved lattice  NRQCD.
The results of  nonperturbative lattice NRQCD simulations are typically given
for  $a\sim 1/(vm_b)$ \cite{Dowdall:2011wh,Dowdall:2013jqa}. The use of
relatively  large values of the  lattice spacing ensures the suppression of
the singular ultraviolet cutoff dependence from the higher order $1/(am_b)^n$
terms,  which are not removed  by the finite order matching  and become
important at $a\sim 1/m_b$. At the same time it results in sizable Coulomb
lattice  artifacts proportional to a power of $\alpha_s am_b\sim 1$.  The
correct treatment of the  artifacts is therefore crucial for the
analysis.

In  actual lattice simulations, the quarkonium bound state  parameters are
extracted from the asymptotic behavior of the quark-antiquark propagator at
large Euclidean time. Neglecting the retardation and long distance
nonperturbative effects, which do not affect the Coulomb artifacts under
consideration, this method should reproduce the properties of the solution of
the Schr\"odinger equation for a given NRQCD Hamiltonian on the spatial
lattice. As we have shown above, the perturbative matching of the  four-quark
operators does not correctly account for the Coulomb artifacts and, for
$a\sim 1/(vm_b)$ and $v\sim\alpha_s$,   results in ${\cal O}(1)$ error in the
prediction for the spectrum. For example, the one-loop matching of the
spin-flip four-quark operator with the spurious linear Coulomb artifact
gives the value of the bottomonium hyperfine splitting \cite{Hammant:2011bt},
which overshoots the predictions of  perturbative QCD \cite{Kniehl:2003ap} by
almost  a factor of two, in clear conflict with the general understanding of
the  heavy quarkonium dynamics.

In practice  the effect of the lattice artifacts is reduced by  numerical
extrapolation of the data to $a=0$ \cite{Dowdall:2011wh,Dowdall:2013jqa}. The
extrapolation below  $a\sim 1/m_b$ in this case is justified because  for the
typical values of lattice spacing the numerical  effect of the $1/(am_b)^n$
terms on the data points is small.  This extrapolation effectively removes
all the lattice artifacts including the $(a\Lambda_{QCD})^n$ terms associated
with the effect of lattice regularization on the dynamics at the confinement
scale $\Lambda_{QCD}$.\footnote{For the bottomonium ground state this
contribution is numerically suppressed with respect to the Coulomb artifacts
since $\Lambda_{QCD}\ll \alpha_s m_q$.} The problem of the perturbative
matching breakdown,  however, is not fully fixed  by this  procedure.  Since
the radiatively improved lattice result is supposed to be free of linear
artifacts, the extrapolation is performed through a constrained fit of the
data points  by a polynomial in $a$ with  {\it vanishing} linear term.  Since
the  bare lattice data are free of the linear Coulomb artifacts the one-loop
perturbative matching in fact {\it introduces} a linear dependence  of the
radiatively improved result on $a$, which leads to  a systematic error of the
fit.  For example, in the analysis of the bottomonium hyperfine splitting
\cite{Dowdall:2013jqa} based on the one-loop  perturbative matching
\cite{Hammant:2011bt,Hammant:2013sca} this  error exceeds  $10\%$ of the
total result and is beyond the estimated uncertainty interval of the
radiatively  improved  ${\cal O}(v^6)$ lattice NRQCD analysis
\cite{Liu:2016fus}. At the same time the perturbative matching  can  be used
for the self-consistent analysis of the quarkonium spectrum within the above
extrapolation scheme if the Coulomb artifacts are removed from the Wilson
coefficients by means of the asymptotic expansion \cite{Baker:2015xma} or a
numerical fit \cite{Liu:2016fus}.

\vspace{10mm}

%\acknowledgements
\noindent
{\bf Acknowledgements}\\[3mm]
We would like to thank Tao Liu for useful discussions and comments on the
manuscript. This work  is supported in part by NSERC. The work of A.P. is
supported in part by the  Perimeter Institute for Theoretical Physics.
%Research at Perimeter Institute is supported by the Government of Canada
%through Industry Canada and by the Province of Ontario through the Ministry
%of Research and Innovation.

%\newpage
%%%%%%%%%%%%%%%%%%%%%%%%%%%%%%%%%%%%%%%%%%%%%%%%%%%%%%%%%%%%

\end{document}